# The Prompting Brain: Neurocognitive Markers of Expertise in Guiding Large Language Models

Hend S. Al-Khalifa
*Information Technology Department, College of Computer and Information Sciences*
*King Saud University*
Riyadh, Saudi Arabia
hendk@ksu.edu.sa

Raneem Almansour
Radiological sciences, College of Applied Medical Sciences
King Saud University
Riyadh, Saudi Arabia
443202782@student.ksu.edu.sa

Layan Abdulrahman Alhuasini
*Radiological sciences, College of Applied Medical Sciences*
*King Saud University*
Riyadh, Saudi Arabia
443202718@student.ksu.edu.sa

Alanood Alsaleh
Radiological sciences, College of Applied Medical Sciences
King Saud University
Riyadh, Saudi Arabia
alalanoud@ksu.edu.sa

Mohamad-Hani Temsah
Pediatric Department, College of Medicine
King Saud University
Riyadh, Saudi Arabia
mtemsah@ksu.edu.sa

Ashwag Rafea S Alruwaili
*Radiological sciences, College of Applied Medical Sciences*
*King Saud University*
Riyadh, Saudi Arabia
ashalruwaili@ksu.edu.sa

## Abstract

Prompt engineering has rapidly emerged as a critical skill for effective interaction with large language models (LLMs). However, the cognitive and neural underpinnings of this expertise remain largely unexplored. This paper presents findings from a cross-sectional pilot fMRI study investigating differences in brain functional connectivity and network activity between experts and intermediate prompt engineers. Our results reveal distinct neural signatures associated with higher prompt engineering literacy, including increased functional connectivity in brain regions such as the left middle temporal gyrus and the left frontal pole, as well as altered power-frequency dynamics in key cognitive networks. These findings offer initial insights into the neurobiological basis of prompt engineering proficiency. We discuss the implications of these neurocognitive markers in Natural Language Processing (NLP). Understanding the neural basis of human expertise in interacting with LLMs can inform the design of more intuitive human-AI interfaces, contribute to cognitive models of LLM interaction, and potentially guide the development of AI systems that better align with human cognitive workflows. This interdisciplinary approach aims to bridge the gap between human cognition and machine intelligence, fostering a deeper understanding of how humans learn and adapt to complex AI systems.

## 1. Introduction

The advent of sophisticated Large Language Models (LLMs) has revolutionized numerous domains, yet their effective utilization hinges significantly on the art and science of prompt engineering. The ability to craft precise and effective prompts that guide LLMs towards the desired outputs is becoming an indispensable skill. While the Natural Language Processing (NLP) community has made substantial strides in developing more capable LLMs and understanding their internal workings [1], [2], the human side of this interaction—specifically, the cognitive and neural mechanisms that differentiate expert prompt engineers from novices—remains a nascent area of research. As LLMs become increasingly integrated into daily tasks and complex problem-solving, understanding the cognitive demands and neural adaptations associated with mastering prompt engineering is crucial for optimizing human-AI collaboration.

This study took an interdisciplinary approach, leveraging neuroimaging techniques to explore the neural correlations of prompt engineering expertise. We hypothesize that expertise in prompt engineering is not merely a matter of accumulated knowledge but is also reflected in distinct patterns of brain function and connectivity. The primary contributions of this work are threefold. First, we introduce a novel methodology for quantifying prompt engineering literacy and use it to differentiate between expert and intermediate users. Second, we provide the first, to our knowledge, empirical evidence from fMRI data identifying distinct neural signatures (including altered functional connectivity in language and executive control regions, and different network power dynamics) associated with prompt engineering expertise. Third, we discuss the potential implications of these neurocognitive markers for the NLP field, particularly for designing more intuitive human-AI interfaces, developing cognitive models of LLM interaction, and guiding the creation of AI systems that better align with human cognitive processes. By identifying these neural markers, we aim to provide a foundational understanding of the cognitive processes that underpin successful human-LLM interaction. This research seeks to bridge insights from cognitive neuroscience with challenges and opportunities within NLP. Understanding how the human brain adapts to the task of "programming" LLMs through natural language can inform the design of more intuitive prompting interfaces, contribute to the development of training programs for prompt engineering, and potentially inspire new architectures for LLMs that are more attuned to human cognitive styles.

Our pilot study employed functional magnetic resonance imaging (fMRI) to compare resting-state brain activity between individuals classified as experts and intermediate prompt engineers based on a validated Prompt Engineering Literacy Scale. We focused on differences in functional connectivity and network power dynamics to identify objective neural indicators of prompting proficiency. The findings presented herein offer preliminary evidence for such neural distinctions, and open avenues for future research at the intersection of neuroscience, cognitive science, and NLP. This work aims to contribute to a more holistic understanding of human-AI interaction, moving beyond model-centric analyses to incorporate the cognitive and neural dimensions of the human user.

## 2. Related Work

With the development of LLMs, the field of prompt engineering has grown rapidly. Initial work focused on empirical discoveries of effective prompting strategies [1], [3]. More recent research has sought to systematize prompt engineering, categorizing techniques and exploring their efficacy across different models and tasks [4], [5]. For instance, Priyadarshana et al. (2024) provide a review of prompt engineering types, methods, and tasks, particularly in the context of digital mental health, highlighting techniques like n-shot prompting and chain-of-thought (CoT) prompting. CoT prompting, which encourages models to produce intermediate reasoning steps, has been shown to improve performance on complex reasoning tasks [3], [6].

While much research has focused on the AI side of prompt engineering, the human cognitive aspects are less understood. Some studies have explored how human interaction styles and cognitive biases affect LLM performance [7], but direct neuroimaging investigations of prompt engineering expertise are scarce. Cognitive science offers theories on expertise development, suggesting that experts develop specialized mental representations and processing strategies [8]. It is plausible that prompt engineering expertise involves similar cognitive adaptations, potentially reflected in neural activity.

Neuroimaging studies of related complex cognitive skills, such as programming or problem-solving, have identified involvement of prefrontal cortex (executive functions, planning), parietal cortex (spatial reasoning, attention), and temporal cortex (language, semantic memory) [9], [10]. For instance, studies on software developers have shown differences in brain activation patterns between experts and novices during code comprehension tasks, often implicating language and working memory networks [9]. Our study builds on this by specifically investigating the neural correlates of expertise in interacting with LLMs through natural language prompts, a unique form of human-computer interaction that blends linguistic and logical reasoning.

The intersection of NLP and cognitive neuroscience is an emerging and rapidly advancing area. Researchers are increasingly exploring how neural data can inform AI models and, conversely, how computational models can provide insights into brain function [11], [12]. Recent work by [13] introduced a unified computational framework connecting acoustic, speech, and word-level linguistic structures to study the neural basis of everyday conversations using electrocorticographic (ECoG) and the Whisper model. They demonstrated an alignment between the model's internal processing hierarchy and the cortical hierarchy for language. Further supporting this line of inquiry, [14] utilized pre-trained NLP models with intracranial recordings to discover neural signals reflecting speech production, comprehension, and their transitions during natural conversation, highlighting broadly distributed frontotemporal activities specific to the words and sentences being conveyed. Complementing these findings, [15] also showed that brain embeddings derived from intracranial recordings in the inferior frontal gyrus (IFG) share common geometric patterns with contextual embeddings from deep language models (DLMs), suggesting a vector-based neural code for natural language processing. Our work, while focusing on the specific skill of prompt engineering expertise rather than general language processing in conversation or direct neural encoding of speech, shares the overarching goal of understanding the neural underpinnings of human interaction with complex language-based systems. The methodologies and findings from these recent studies, particularly the emphasis on aligning computational model representations (embeddings) with neural activity, provide a valuable context for interpreting our results on how expertise in manipulating such systems might manifest neurally. Our work contributes to this interdisciplinary dialogue by examining the neural underpinnings of a critical skill in the current NLP landscape, potentially offering insights into both human-centered AI design and cognitive theories of expertise in technologically mediated tasks.

# 3. Methods

## 3.1 Study Design and Participants

This research was conducted as a cross-sectional pilot study. A total of 22 participants (aged 18–45 years) were recruited. Participants were screened to meet specific inclusion criteria, including right-handedness and no history of neurological or psychiatric disorders. Table 1 shows participants demographics. Ethical approval was obtained from our institutional review board (No. E-24-9142), and all participants provided informed consent prior to their involvement in the study.

Participants were classified into two groups, "intermediate" and "expert," based on their scores on a custom-developed instrument called Prompt Engineering Literacy Scale (PELS) (see Appendix A). This scale consisted of 10 questions, each rated on a 5-point Likert scale, yielding a maximum possible score of 50. Individuals scoring above 37 were categorized as experts, while those scoring 37 (70%) or below were categorized as intermediate. The scale was designed to assess four key dimensions of prompting expertise: (1) Prompt Construction and Clarity, (2) Advanced Prompting Techniques (knowledge of various prompting techniques), (3) Verification and Optimization (methods to validate and optimize the quality of AI responses), and (4) Ethical and Cultural Sensitivity in prompt formulation. The PELS instrument was developed through a rigorous process involving expert consultation, literature review and pilot testing. Initial items were generated based on a comprehensive review of prompt engineering literature [16] and consultation with three prompt engineering experts with over two years of experience in the field.

To stratify participants into "expert" and "intermediate" groups, we applied a threshold of 37 out of 50 on the PELS. This cutoff was informed by a combination of empirical score distribution and expert consensus, consistent with standard practices in early-stage scale deployment [17], [18]. During pilot validation, individuals with demonstrated prompting proficiency (e.g., graduate students with AI coursework or industry experience) consistently scored above this threshold. In addition, qualitative feedback from participants during pilot testing supported the idea that scores above this level better reflected advanced prompting fluency.

While this cutoff is exploratory rather than norm-referenced, it functioned as a practical criterion for distinguishing proficiency levels in a small sample. This approach aligns with preliminary classification practices in early psychometric work [19]. Future versions of the PELS could be refined using item response theory [20] or receiver operating characteristic (ROC) analysis [21] to establish statistically optimized thresholds.

To ensure the validity and reliability of the instrument, we conducted both external and internal consistency assessments. Before the main study, external validity was established through expert review by a panel of experts in AI and LLMs as well as neuroscientists, who evaluated the instrument for content relevance and clarity. Their feedback was incorporated into the final version. The questionnaire was subjected to a pilot study for internal consistency with a random sample of subjects to assess its validity and reliability. The pilot study yielded a Cronbach's alpha of ($\alpha$ = 0.90), indicating high reliability and validity of the instrument.

**Table 1: Participants' Demographics**

| Characteristics | Intermediate | Expert |
|---|---|---|
| **Number of users** | 10 | 12 |
| **Gender ratio (female: male)** | 8:2 Females=8 Males= 2 | 7:5 Females=7 Males =5 |
| **Age (Mean±SD)** | 22.5± 4.8 | 22.25± 3.8 |
| **Education Level** | Bachelor (9) Ph.D. (1) | Bachelor (11) Ph.D. (1) |
| **PELS scores (Mean±SD)** | 32.2 ± 3.3 | 40.9± 3.42 |

## 3.2 Data Acquisition

Magnetic Resonance Imaging (MRI) data were acquired using a Siemens MAGNETOM Spectra (DE) 3T scanner equipped with a standard head coil. During the fMRI scan, participants were instructed to relax, remain still, keep their eyes open, and stay awake, consistent with a resting-state fMRI paradigm. High-resolution T1-weighted anatomical images were also acquired for registration purposes. Table 2 details the specific parameters used for both the T1-weighted anatomical scans and the resting-state fMRI.

**Table 2: MR Imaging parameters used for both the T1-weighted anatomical scans and the resting-state fMRI BOLD EPI sequence, which are: Repetition Time (TR), Echo Time (TE), flip angle, field of view (FOV), voxel size, matrix size, and time.**

|  | TR | TE | Flip angle | FOV | Voxel size | Matrix | Time |
|---|---|---|---|---|---|---|---|
| **T1-weighted-(3D)** | 1900ms | 2.42ms | 9 degrees | 250mm | 1.0 ×1.0 ×1.0mm | 250*250 | 5.11 min |
| **Resting-fMRI** | 2270ms | 27ms | 90 degrees | 250 mm | 2.7×2.7×2.5mm | 93*93 | 11.2 min |

| gradient-echo | | | | | | |

## 3.3 fMRI Data Analysis

Functional MRI data were preprocessed and analyzed using two complementary pipelines to ensure robust identification of group-level differences in brain network activity and functional integration.

**Preprocessing:** Standard fMRI preprocessing steps were applied, including motion correction, slice-timing correction, spatial normalization to a standard template (e.g., MNI space), and spatial smoothing. Nuisance regression was performed to remove effects of motion parameters, white matter signal, and cerebrospinal fluid signal.

**Independent Component Analysis (ICA):** Group ICA was performed using the GIFT (Group ICA of fMRI Toolbox) software. This data-driven approach decomposes the fMRI data into a set of spatially independent components and their corresponding time courses. We focused on identifying differences in the spectral power of these components between the expert and intermediate groups, particularly the ratio of low-frequency power (LF, e.g. 0.01-0.08 Hz) to high-frequency power (HF, e.g. >0.1 Hz) within established resting-state networks. The number of components (n=20) was chosen a priori based on previous resting-state fMRI studies of expertise.

**Seed-to-Voxel Connectivity Analysis**: Seed-based functional connectivity analysis was conducted using the CONN toolbox. Regions of interest (ROIs) identified from prior literature or ICA results were used as seeds. For each participant, Pearson correlation coefficients were calculated between the mean time series of each seed ROI and the time series of all other voxels in the brain. These correlation maps were then converted to z-scores using Fisher's r-to-z transformation to allow for group-level statistical comparisons (two-sample t-tests) between the expert and intermediate groups, controlling for relevant covariates if necessary.

## 4. Results

The analysis of fMRI data revealed significant differences between the expert and intermediate prompt engineering groups, both in terms of intrinsic network activity and specific functional connections.

## 4.1 Differences in Network Power Ratios

Using separate ICA analyses for each group, Power_LF/Power_HF ratios were computed to assess low-frequency dominance. Three key components showed higher ratios in the expert group, indicating greater low-frequency synchronization: Experts demonstrated strikingly higher

Low-to-High Frequency Power Ratios (LHR) in core cognitive networks, as shown in Table 3 and Figure 1.

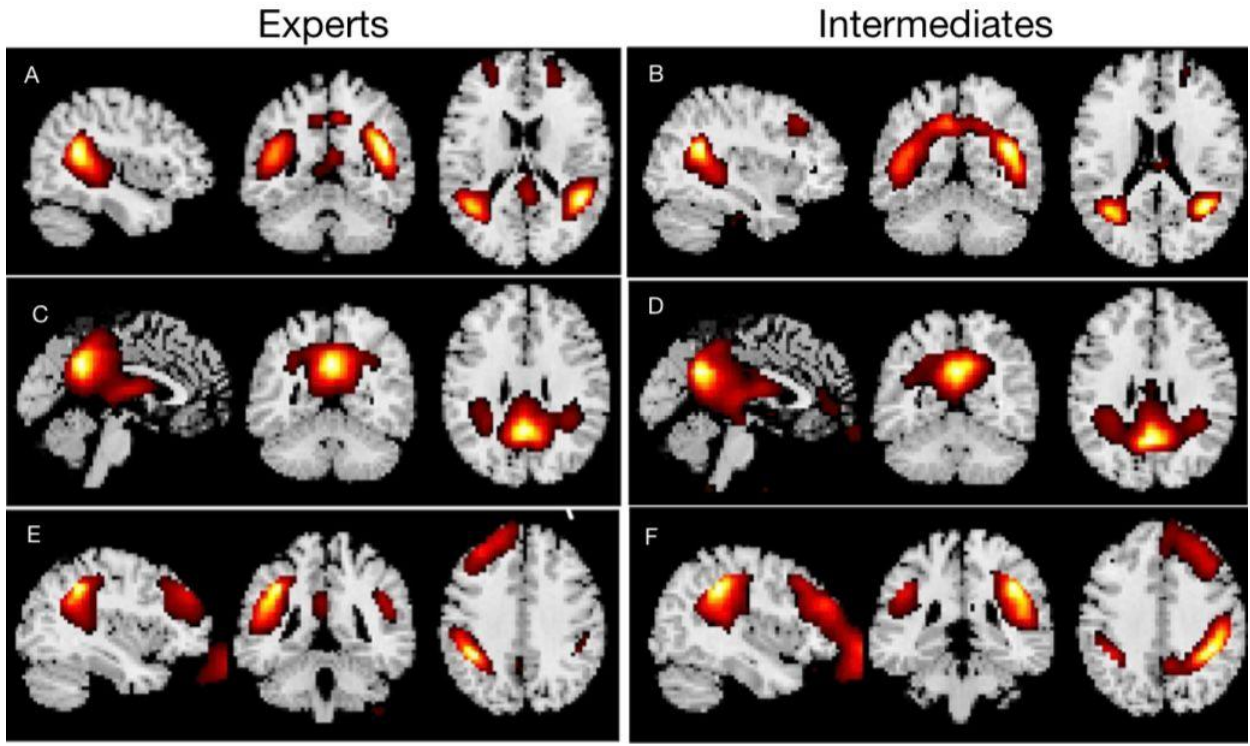

Figure 1: Experts showed higher Power_LF/Power_HF ratios in key networks: Ventral Visual Network (VVN) (example: A,B) – 63.0 vs. 36.7, Posterior Default Mode Network (pDMN) (example: C,D)– 44.4 vs. 33.2, and Left Lateral Parietal Network (LLPN) (example; E,D)– 53.3 vs. 36.7. These differences suggest stronger low-frequency synchronization in experts. The LLPN is part of the default mode network and is associated with semantic processing and episodic memory retrieval.

Table 3: Group Differences in Low-to-High Frequency Power Ratios (Power_LF/Power_HF)

| Network | Power_LF/Power_HF (Experts) | Power_LF/Power_HF (Intermediates) |
|---|---|---|
| Ventral Visual Network (VVN) | 63 | 36.7 |
| Left Lateral Parietal Network | 53.3 | 36.7 |

| Posterior Default Mode Network | 44.4 | 33.2 |

As indicated by the preliminary data (Figure 1 and Table 3), these differences were prominent in:
- The **Ventral Visual Network (VVN)**: Experts showed a ratio of approximately 63.0 versus 36.7 in intermediates.
- The **Posterior Default Mode Network (pDMN)**: Experts showed a ratio of approximately 44.4 versus 33.2 in intermediates.
- The Left Lateral Parietal Network (LLPN) (potentially referring to a component of the fronto-parietal network or language network): Experts showed a ratio of approximately 53.3 versus 36.7 in intermediates. The LLPN is part of the default mode network and is associated with semantic processing and episodic memory retrieval.

These differences suggest stronger low-frequency synchronization and potentially more stable intrinsic network dynamics in individuals with higher prompt engineering expertise.

## 4.2 Differences in Functional Connectivity

Seed-to-voxel connectivity analyses identified specific brain regions where functional connectivity differed significantly between the two groups:

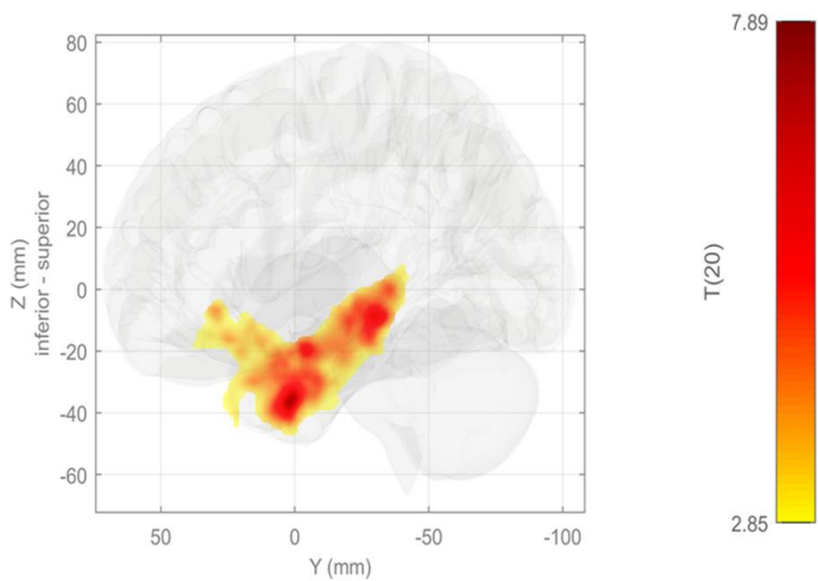

**Figure 2: Increased connectivity in the left middle temporal gyrus (p < 0.03) in expert group**

- **Left Middle Temporal Gyrus (MTG):** The expert group demonstrated significantly increased functional connectivity involving the left MTG (p < 0.03), as suggested by Figure 2. The specific seed or target regions connected to the left MTG would be detailed based on the full analysis, but this region is classically involved in language processing, semantic memory, and multimodal integration.

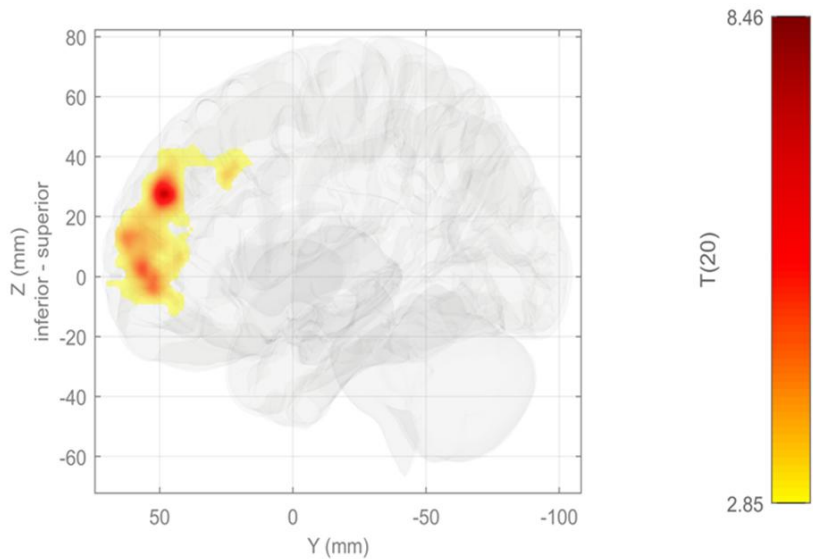

**Figure 3: Increased connectivity in the left frontal pole (p < 0.05) in expert group**

- **Left Frontal Pole (FP):** Similarly, the expert group showed increased functional connectivity in the left frontal pole (p < 0.05), as suggested by the placeholder for Figure 3. The frontal pole is associated with higher-order cognitive functions, including planning, goal-directed behavior, abstract reasoning, and metacognition.

To validate these results, Fractional Amplitude of Low-Frequency Fluctuations (fALFF), where fALFF is defined as the ratio of the ALFF at each voxel divided by the signal power over the entire frequency range values were extracted from a unified ICA of all 22 participants. Experts demonstrated consistently lower fALFF across multiple components, reflecting reduced spontaneous fluctuations and potentially more efficient neural processing at rest, consistent with the cognitive control role of fALFF (Figure 4).

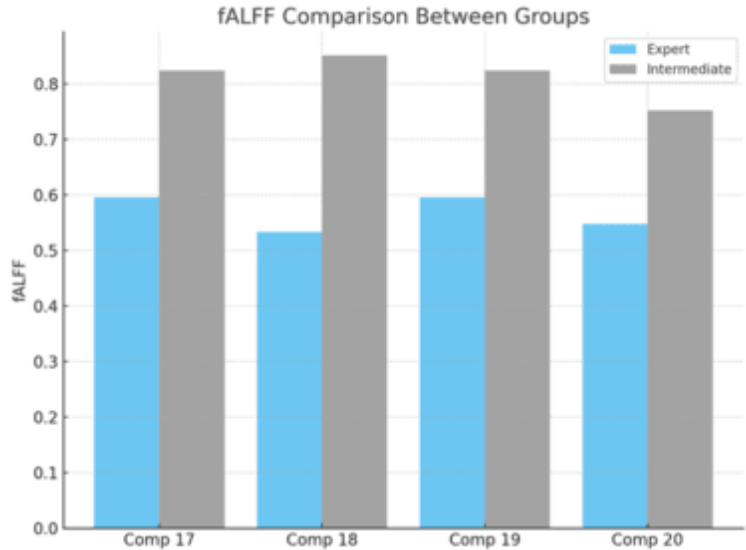

**Figure 4: Experts showed consistently lower fALFF values across key components: 0.534 vs. 0.852 in Component 18, 0.596 vs. 0.824 in Components 17 and 19, and 0.548 vs. 0.752 in Component 20, indicating reduced spontaneous low-frequency activity in experts.**

The components shown in Figure 4 (Components 18-20) correspond to specific brain networks: Component 18 is responsible for spatial attention, eye movement planning, Component 19 is part of the medial visual network, and Component 20 is a region involved in cognitive control network. While we interpret lower fALFF values in experts as potentially reflecting more efficient neural processing, several alternative explanations warrant consideration. Lower fALFF could also indicate differences in neurovascular coupling, reflect compensatory mechanisms, or represent strategy-specific adaptations rather than general efficiency. Additionally, amplitude differences can be influenced by various physiological factors including cerebral blood flow, metabolism, and neurotransmitter dynamics [22]. The consistency of this pattern across multiple networks, however, suggests a systematic difference in neural dynamics between expert and intermediate prompt engineers that merits further investigation with task-based paradigms.

## 5. Discussion

The findings of this pilot study provide initial neurobiological evidence distinguishing expert prompt engineers from those with intermediate skills, with specific, actionable implications for both NLP and HCI design. The observed differences in low-frequency power ratios within key brain networks, such as the visual network (VVN), posterior default mode network (pDMN), and left lateral parietal network (LLPN) suggest that expertise in prompt engineering may be associated with more organized and stable intrinsic neural activity. Higher LF/HF power ratios are often interpreted as reflecting more efficient neural processing or greater network integration [23], [24] indicate that expert prompt engineers may possess neural systems more finely tuned for managing the cognitive demands of prompting. For example, the involvement of the DMN may support internal thought processes and semantic integration crucial for prompt formulation [25], while visual and parietal network activity may underlie the conceptual and spatial structuring of prompt elements [26].

The increased functional connectivity in the left middle temporal gyrus (MTG) and left frontal pole among experts is particularly notable. The MTG, a hub for semantic processing and contextual integration [27], [28], likely supports experts' refined ability to access and manipulate language representations. The frontal pole, implicated in strategic planning and cognitive control [29], [30], may enable experts to construct and iteratively refine prompts using higher-order strategies. These insights suggest that reducing cognitive load in this region could benefit novices, a goal that might be achieved through interface designs incorporating structured planning scaffolds, such as prompt templates with predefined sections for context, constraints, and examples.

Although this study focused on resting-state connectivity, prior research demonstrates strong links between resting-state network architecture and performance on complex tasks [31], [32], [33]. Our findings, interpreted through this lens, imply that the neural systems supporting expert prompt engineering are primed for efficient engagement during task performance, even in the absence of explicit prompts. Notably, the involvement of visual networks, while seemingly counterintuitive for a linguistic task, aligns with theories of visual mental imagery and mental model construction. Experts may simulate AI responses and mentally visualize prompt-output relationships, engaging visual processing regions to spatially organize abstract linguistic structures [9], [34], [35]. These strategies could be supported in HCI through visual interfaces that represent prompt logic and flow, such as flowchart-style diagrams, offloading mental simulation and aligning with expert cognitive styles.

Our results also resonate with broader trends in computational neuroscience and NLP that explore the alignment between model embeddings and neural representations. For instance, Goldstein et al. [13], [15] demonstrate how unified embedding spaces can map onto cortical hierarchies, while Cai et al. [14] highlight how language model activity parallels distributed neural dynamics during speech comprehension. Though our focus is specific to prompt engineering expertise, the underlying principle remains the same: complex language interactions, like those required to prompt LLMs effectively, are grounded in specialized configurations of brain networks. The enhanced connectivity we observe in the left MTG and frontal pole may reflect a domain-specific neural adaptation that facilitates effective engagement with the abstract embedding spaces used by LLMs suggesting that future models could be designed to anticipate or support these expert cognitive strategies.

In summary, these neurocognitive findings provide a foundation for cognitively-informed design principles in NLP and HCI. Interfaces tailored to support semantic retrieval, visual reasoning, and executive planning could ease the transition from novice to expert. Educational tools might train users to progressively engage in the relevant neural systems, while prompt engineering platforms could incorporate adaptive scaffolds that evolve with user proficiency. Moreover, neural markers of expertise could serve as user-centered evaluation metrics, favoring tools that reduce unnecessary executive demand while preserving task performance. Ultimately, by integrating neuroscientific insights with interface design and model development, we can foster more intuitive, efficient, and human-aligned interactions with AI systems.

# 6. Conclusion

This pilot fMRI study provides novel preliminary evidence for distinct neural signatures associated with prompt engineering expertise. We observed that individuals with higher proficiency in prompt engineering exhibit altered low-frequency power dynamics in key cognitive networks (VVN, pDMN, LLPN) and increased functional connectivity in brain regions critical for language processing and higher-order cognition (left MTG, left frontal pole). These findings, while requiring replication and extension, offer a first glimpse into the neurobiological underpinnings of this increasingly vital skill.

The implications of this research extend into the domain of Natural Language Processing, particularly in the context of human-AI interaction. By understanding the neural basis of how humans effectively interact with and guide LLMs, we can pave the way for more intuitive AI interfaces, develop more targeted training methodologies for prompt engineering, and potentially inform the design of next-generation LLMs that are better aligned with human cognitive architectures. Future research should aim to expand on these findings, perhaps by investigating neural changes during active prompting tasks or longitudinally tracking the development of these neural markers as individuals gain prompting expertise. Such interdisciplinary endeavors are crucial for fostering a synergistic relationship between human intelligence and artificial intelligence.

This study has several limitations that should be considered when interpreting the findings. Firstly, the pilot nature of the study means our sample size (N=22) is relatively small. While sufficient for detecting initial group differences, larger cohorts are needed to confirm these results and to explore more subtle effects or individual variability. Secondly, the cross-sectional design does not allow for causal inferences; we can observe associations between expertise and neural patterns, but not whether these neural patterns are a cause or a consequence of developing expertise. Longitudinal studies tracking individuals as they develop prompt engineering skills would be beneficial.

Another limitation concerns the selection of the cutoff score of 37 on the PELS to define expert versus intermediate groups. While the decision was informed by expert review and observed clustering in pilot data, it remains a preliminary, heuristic threshold. The scale has not yet undergone advanced psychometric calibration such as latent class analysis or IRT modeling, which are considered best practices in modern test development [20]. This limitation highlights the need for future larger-scale validation studies to establish stable, generalizable classification boundaries. As the field of prompt engineering literacy matures, more robust data-driven techniques can be employed to refine and validate such thresholds [17]. Additionally, our reliance on resting-state fMRI alone limits our ability to directly link neural patterns to specific cognitive processes involved in prompt engineering. Complementary task-based fMRI studies where participants actively engage in prompt formulation and refinement would provide more direct evidence of the neural systems supporting this skill. Such studies could employ parametric designs manipulating prompt complexity or constraints to isolate specific cognitive components of prompt engineering expertise.

Finally, while our findings suggest neural markers associated with prompt engineering expertise, the rapidly evolving nature of AI systems means that the cognitive demands of prompt engineering may change over time. As large language models and other AI systems continue to advance, the specific skills and strategies that constitute effective prompt engineering may shift, potentially altering the associated neural correlates. Longitudinal research tracking these changes would provide valuable insights into the co-evolution of human cognitive adaptations and AI capabilities.

# References

[1]  T. Brown *et al.*, "Language Models are Few-Shot Learners," in *Advances in Neural Information Processing Systems*, Curran Associates, Inc., 2020, pp. 1877–1901. Accessed: May 15, 2025. [Online]. Available: https://papers.nips.cc/paper/2020/hash/1457c0d6bfcb4967418bfb8ac142f64a-Abstract.html

[2]  A. Vaswani *et al.*, "Attention is All you Need," in *Advances in Neural Information Processing Systems*, Curran Associates, Inc., 2017. Accessed: May 13, 2025. [Online]. Available: https://proceedings.neurips.cc/paper/2017/hash/3f5ee243547dee91fbd053c1c4a845aa-Abstract.html

[3]  J. Wei *et al.*, "Chain-of-thought prompting elicits reasoning in large language models," in *Proceedings of the 36th International Conference on Neural Information Processing Systems*, in NIPS '22. Red Hook, NY, USA: Curran Associates Inc., Nov. 2022, pp. 24824–24837.

[4]  Y. H. P. P. Priyadarshana, A. Senanayake, Z. Liang, and I. Piumarta, "Prompt engineering for digital mental health: a short review," *Front. Digit. Health*, vol. 6, Jun. 2024, doi: 10.3389/fdgth.2024.1410947.

[5]  J. White *et al.*, "A Prompt Pattern Catalog to Enhance Prompt Engineering with ChatGPT," in *Proceedings of the 30th Conference on Pattern Languages of Programs*, in PLoP '23. USA: The Hillside Group, Oct. 2023, pp. 1–31.

[6]  T. Kojima, S. S. Gu, M. Reid, Y. Matsuo, and Y. Iwasawa, "Large language models are zero-shot reasoners," in *Proceedings of the 36th International Conference on Neural Information Processing Systems*, in NIPS '22. Red Hook, NY, USA: Curran Associates Inc., Nov. 2022, pp. 22199–22213.

[7]  S. Malberg, R. Poletukhin, C. Schuster, and G. G. Groh, "A Comprehensive Evaluation of Cognitive Biases in LLMs," in *Proceedings of the 5th International Conference on Natural Language Processing for Digital Humanities*, M. Hämäläinen, E. Öhman, Y. Bizzoni, S. Miyagawa, and K. Alnajjar, Eds., Albuquerque, USA: Association for Computational Linguistics, May 2025, pp. 578–613. Accessed: May 16, 2025. [Online]. Available: https://aclanthology.org/2025.nlp4dh-1.50/

[8]  K. A. Ericsson and W. Kintsch, "Long-term working memory," *Psychol Rev*, vol. 102, no. 2, pp. 211–245, Apr. 1995, doi: 10.1037/0033-295x.102.2.211.

[9]  B. Floyd, T. Santander, and W. Weimer, "Decoding the Representation of Code in the Brain: An fMRI Study of Code Review and Expertise," in *2017 IEEE/ACM 39th International Conference on Software Engineering (ICSE)*, May 2017, pp. 175–186. doi: 10.1109/ICSE.2017.24.

[10] J. Siegmund, C. Kästner, J. Liebig, S. Apel, and S. Hanenberg, "Measuring and modeling programming experience," *Empirical Softw. Engg.*, vol. 19, no. 5, pp. 1299–1334, Oct. 2014, doi: 10.1007/s10664-013-9286-4.
[11] C. Caucheteux and J.-R. King, "Brains and algorithms partially converge in natural language processing," *Commun Biol*, vol. 5, no. 1, pp. 1–10, Feb. 2022, doi: 10.1038/s42003-022-03036-1.
[12] N. Kriegeskorte and P. K. Douglas, "Interpreting encoding and decoding models," *Current Opinion in Neurobiology*, vol. 55, pp. 167–179, Apr. 2019, doi: 10.1016/j.conb.2019.04.002.
[13] A. Goldstein *et al.*, "A unified acoustic-to-speech-to-language embedding space captures the neural basis of natural language processing in everyday conversations," *Nat Hum Behav*, pp. 1–15, Mar. 2025, doi: 10.1038/s41562-025-02105-9.
[14] J. Cai *et al.*, "Natural language processing models reveal neural dynamics of human conversation," *Nat Commun*, vol. 16, no. 1, p. 3376, Apr. 2025, doi: 10.1038/s41467-025-58620-w.
[15] A. Goldstein *et al.*, "Alignment of brain embeddings and artificial contextual embeddings in natural language points to common geometric patterns," *Nat Commun*, vol. 15, no. 1, p. 2768, Mar. 2024, doi: 10.1038/s41467-024-46631-y.
[16] D. Federiakin, D. Molerov, O. Zlatkin-Troitschanskaia, and A. Maur, "Prompt engineering as a new 21st century skill," *Front. Educ.*, vol. 9, Nov. 2024, doi: 10.3389/feduc.2024.1366434.
[17] G. O. Boateng, T. B. Neilands, E. A. Frongillo, H. R. Melgar-Quiñonez, and S. L. Young, "Best Practices for Developing and Validating Scales for Health, Social, and Behavioral Research: A Primer," *Front. Public Health*, vol. 6, Jun. 2018, doi: 10.3389/fpubh.2018.00149.
[18] R. F. DeVellis, *Scale Development: Theory and Applications*. Thousand Oaks, Calif.: SAGE Publications, Inc, 2012.
[19] L. A. Clark and D. Watson, *Constructing validity: Basic issues in objective scale development*. in Methodological issues and strategies in clinical research, 4th ed. Washington, DC, US: American Psychological Association, 2016, p. 203. doi: 10.1037/14805-012.
[20] S. E. Embretson and S. P. Reise, *Item Response Theory for Psychologists*. New York: Psychology Press, 2013. doi: 10.4324/9781410605269.
[21] C. E. Metz, "Basic principles of ROC analysis," *Semin Nucl Med*, vol. 8, no. 4, pp. 283–298, Oct. 1978, doi: 10.1016/s0001-2998(78)80014-2.
[22] Q.-H. Zou *et al.*, "An improved approach to detection of amplitude of low-frequency fluctuation (ALFF) for resting-state fMRI: fractional ALFF," *J Neurosci Methods*, vol. 172, no. 1, pp. 137–141, Jul. 2008, doi: 10.1016/j.jneumeth.2008.04.012.
[23] G. Buzsáki and A. Draguhn, "Neuronal oscillations in cortical networks," *Science*, vol. 304, no. 5679, pp. 1926–1929, Jun. 2004, doi: 10.1126/science.1099745.
[24] N. K. Logothetis, J. Pauls, M. Augath, T. Trinath, and A. Oeltermann, "Neurophysiological investigation of the basis of the fMRI signal," *Nature*, vol. 412, no. 6843, pp. 150–157, Jul. 2001, doi: 10.1038/35084005.
[25] R. L. Buckner, J. R. Andrews-Hanna, and D. L. Schacter, "The Brain's Default Network," *Annals of the New York Academy of Sciences*, vol. 1124, no. 1, pp. 1–38, 2008, doi: 10.1196/annals.1440.011.
[26] M. Corbetta and G. L. Shulman, "Control of goal-directed and stimulus-driven attention in the brain," *Nat Rev Neurosci*, vol. 3, no. 3, pp. 201–215, Mar. 2002, doi: 10.1038/nrn755.
[27] J. R. Binder, R. H. Desai, W. W. Graves, and L. L. Conant, "Where is the semantic system? A critical review and meta-analysis of 120 functional neuroimaging studies," *Cereb Cortex*, vol. 19, no. 12, pp. 2767–2796, Dec. 2009, doi: 10.1093/cercor/bhp055.

[28] G. Hickok and D. Poeppel, "The cortical organization of speech processing," *Nat Rev Neurosci*, vol. 8, no. 5, pp. 393–402, May 2007, doi: 10.1038/nrn2113.

[29] D. Badre and M. D'Esposito, "Is the rostro-caudal axis of the frontal lobe hierarchical?," *Nat Rev Neurosci*, vol. 10, no. 9, pp. 659–669, Sep. 2009, doi: 10.1038/nrn2667.

[30] E. Koechlin and C. Summerfield, "An information theoretical approach to prefrontal executive function," *Trends in Cognitive Sciences*, vol. 11, no. 6, pp. 229–235, Jun. 2007, doi: 10.1016/j.tics.2007.04.005.

[31] I. Tavor, O. Parker Jones, R. B. Mars, S. M. Smith, T. E. Behrens, and S. Jbabdi, "Task-free MRI predicts individual differences in brain activity during task performance," *Science*, vol. 352, no. 6282, pp. 216–220, Apr. 2016, doi: 10.1126/science.aad8127.

[32] M. W. Cole, D. S. Bassett, J. D. Power, T. S. Braver, and S. E. Petersen, "Intrinsic and task-evoked network architectures of the human brain," *Neuron*, vol. 83, no. 1, pp. 238–251, Jul. 2014, doi: 10.1016/j.neuron.2014.05.014.

[33] N. U. F. Dosenbach *et al.*, "Prediction of individual brain maturity using fMRI," *Science*, vol. 329, no. 5997, pp. 1358–1361, Sep. 2010, doi: 10.1126/science.1194144.

[34] J. Pearson, T. Naselaris, E. A. Holmes, and S. M. Kosslyn, "Mental Imagery: Functional Mechanisms and Clinical Applications," *Trends Cogn Sci*, vol. 19, no. 10, pp. 590–602, Oct. 2015, doi: 10.1016/j.tics.2015.08.003.

[35] S. M. Kosslyn, W. L. Thompson, and G. Ganis, *The Case for Mental Imagery*. Oxford University Press.